\documentclass[fleqn,usenatbib]{mnras}

\usepackage{newtxtext,newtxmath}

\usepackage[T1]{fontenc}

\DeclareRobustCommand{\VAN}[3]{#2}
\let\VANthebibliography\thebibliography
\def\thebibliography{\DeclareRobustCommand{\VAN}[3]{##3}\VANthebibliography}

\usepackage{graphicx}	
\usepackage{amsmath}	

\title[Cosmology with a negative $\Lambda$]{Do cosmological observations allow a negative $\Lambda$?}

\author[A. A. Sen et al.]{
Anjan A. Sen,$^{1,2}$\thanks{aasen@jmi.ac.in}
Shahnawaz A. Adil,$^{3}$
Somasri Sen$^{3}$
\\
$^{1}$School of Arts and Sciences, Ahmedabad University, Ahmedabad, Gujarat, India\\
$^{2}$Center for Theoretical Physics, Jamia Millia Islamia, Delhi-110025, India\\
$^{3}$Department of Physics, Jamia Millia Islamia, Delhi-110025, India 
}

\date{Accepted XXX. Received YYY; in original form ZZZ}

\pubyear{2022}

\begin{document}
\label{firstpage}
\pagerange{\pageref{firstpage}--\pageref{lastpage}}
\maketitle

\begin{abstract}
In view of the recent measurement of $H_{0}$ from HST and SH0ES team, we explore the possibility of existence of a negative cosmological constant (AdS vacua in the dark energy sector) in the Universe. In this regard, we consider quintessence fields on top of a negative cosmological constant and compare such construction with $\Lambda$CDM model using a different combination of CMB, SnIa, BAO and $H_{0}$ data. Various model comparison estimators show that quintessence models with a negative $\Lambda$ is either preferred over $\Lambda$CDM or performs equally as $\Lambda$CDM model. This suggests that the presence of a negative $\Lambda$ (AdS ground state) in our Universe, which can naturally arise in string theory, is consistent with cosmological observations. 
\end{abstract}




\section{Introduction}

Our Universe is full of surprises and each surprise updates our understanding of the Universe in a remarkable way. Around two decades ago, one such surprising observation involving SnIa confirmed the accelerated cosmological expansion of our Universe (\cite{Riess:1998cb,Perlmutter:1998np}). This  observational result completely changed the way we believe gravity works. It established that gravity is repulsive on large cosmological scales unlike its attractive nature as Newton and Einstein taught us. To account for this repulsive behaviour of gravity, we need to either include an exotic unseen component called ``Dark Energy" (DE) having negative pressure, in the energy budget of the Universe or need to modify the gravity behaviour on large distance scales (\cite{de1,de2,de3,Copeland:2006wr,Tsujikawa:2013fta,modified}). Both  these approaches to model repulsive gravity which in turn explain the observed accelerated expansion of the Universe, demand new insights to our field theoretic understanding of particle physics as well as gravity.

If DE is the answer to this puzzle, we should know what is the origin of this DE that has negative pressure and is also dark i.e, it does not emit any light. Quantum field theory (QFT) provides a simple candidate for this: the ground state energy or ``Vacuum Energy'' of the fields. Its energy momentum tensor has the structure $\rho_{de} = -p_{de} = \Lambda$ (\cite{vacuum}) which naturally results negative pressure. Do note that for accelerated expansion, $\Lambda > 0$. This is termed as ``Cosmological Constant". Unfortunately such a $\Lambda$ can not be the DE that we observe today as QFT gives a value for $\Lambda$ which is inconsistent with what we need, to explain the observed accelerated expansion (\cite{de1}). Nevertheless, due to its simple structure, $\Lambda$ has always been preferred by cosmologists as a candidate for DE. Together with Cold Dark matter (CDM), they construct the $\Lambda$CDM model in the Friedman-Robertson-Walker (FRW) framework to describe our Universe on large cosmological scales which can account for the accelerated expansion of the Universe. Although theoretically problematic, interestingly the $\Lambda$CDM model matches with almost all the observational data including very precise measurement of Cosmic Microwave Background Radiation (CMB) by Planck's team (\cite{Aghanim:2018eyx}), high redshift SnIa observations (\cite{sn}), and Baryon Acoustic Oscillation (BAO) (\cite{bao}) measurements by large galaxy surveys.

The observational evidence for a positive $\Lambda$ confirms the presence of a scalar field at the non zero positive minimum of its potential. This in turn results the presence of a stable or meta stable de-Sitter (dS) vacuum depending on the nature of the minimum of the potential. It can also be possible that the scalar field is away from the minimum of the potential and is slowly rolling at the positive part of the potential. This scenario gives rise to a DE component, which unlike $\Lambda$, is not a constant but evolves with time giving rise to a varying equation of state $w_{de} = p_{de}/\rho_{de}$ for the DE (for $\Lambda, w_{de} = -1$). Such evolving DE component is termed as ``Quintessence''(\cite{Copeland:2006wr, Tsujikawa:2013fta}). Most of the  quintessence scalar fields, studied in the context of DE, have a dS ground state. Hence whether DE is $\Lambda$ or an evolving quintessence field, asymptotically the Universe always settles in a dS phase in future. 

To construct such a scenario for quintessence field with dS vacuum is extremely challenging from a string theory perspective (\cite{string1, string2, string3, string4}). In fact, it has been argued that string theory might not contain any dS vacuum at all (\cite{swamp1, swamp2, swamp3}). This can have far reaching cosmological implications, in particular for both early (\cite{2019JHEP...11..075G}) and late time accelerating Universe. On the other hand, a slow rolling quintessence field with anti de-Sitter (AdS)  vacuum is theoretically consistent with the current late time acceleration and at the same time, can be naturally obtained in string theory context using symmetry consideration like AdS/CFT correspondence (\cite{adscft}). Unlike the case for a dS vacuum, there are a large number of consistent solutions in AdS background in string theory. So a quintessence field with AdS vacuum is theoretically quite natural. Yet, cosmological studies, in particular the late time cosmology, have mainly focused on the existence of an asymptotically dS phase where an evolving DE is always modelled with a slow-rolling scalar field having either a zero minimum or a  dS vacuum.

Just recently, we have another very interesting cosmological result. This time, it is not the slope of the expansion rate of the Universe as measured in late nineties, but the expansion rate itself at present (at $z=0$). We denote it by $H_{0}$, the Hubble parameter at present. The recent local measurement of $H_{0}$ by SH0ES team using the Cepheids variables in the host SnIa  is $H_{0} = (73.30\pm 1.04)$ Km/sec/Mpc (R21)(\cite{Riess:2021arx}). This is, till date, the most accurate direct model independent measurement of $H_{0}$ using low redshift observations. There are other low redshift observations that also give similar higher values for $H_{0}$ (\cite{Wong:2019kwg,Freedman:2019jwv,Pesce:2020xfe}). The problem arises when we compare R21 result with the indirect high redshift measurement of $H_{0}$ by Planck using CMBR observation with the assumption that the Universe is governed by the $\Lambda$CDM model (\cite{Aghanim:2018eyx}) which gives $H_{0} = (67.36 \pm 0.54)$ Km/sec/Mpc. For the first time, these two measurements for $H_{0}$ are in tensions at a statistical level of $\sim  5\sigma$ This is also called ``Hubble Tension" in the literature (\cite{Verde:2019ivm}). Although $\Lambda$CDM is already theoretically problematic as discussed above, observationally for the first time, $\Lambda$CDM model faces serious challenges due to this huge tension between low and high redshift measurements of the $H_{0}$ parameter and has opened up renewed interests for models beyond $\Lambda$CDM.

Recently a large number models have been proposed to address the ``Hubble Tension" (\cite{DiValentino:2016hlg, DiValentino:2019jae, Vagnozzi:2019ezj, Alestas:2020mvb, Anchordoqui:2019amx, DiValentino:2019exe,2021PhRvD.103h1305B}). Both early Universe (\cite{tanvi, early1,Niedermann:2021vgd, Gomez-Valent:2021cbe, Jiang:2021bab, Karwal:2021vpk, Freese:2021rjq,Cai:2021wgv,2021PhRvD.104f3524V}) as well as late Universe solutions (\cite{late1, late2, late3, late4, late5, late6, late7, late8, late9, late10}) have been put forward in different studies (for a detail review,please see (\cite{reviewh0, Schoneberg:2021qvd, Perivolaropoulos:2021jda})). Interestingly, the quintessence models with AdS vacuum got certain momentum under such efforts (\cite{ads1, ads2, ads3, ads4,2020PhRvD.101h3507Y,2021arXiv210804246Y}). None of these studies ruled out the presence of a negative $\Lambda$, although in statistical term, there is no positive evidence for it in comparison to $\Lambda$CDM model. Also the full CMB likelihood was not considered in any of these studies related to models with AdS vacuum.

In this work, we consider the full CMB likelihood including the Lensing likelihood as given by Planck-2018 data release together with the Pantheon data, BAO measurements as well as the $H_{0}$ measurement by R21. We also consider a varying DE equation state in addition to a constant equation of state for DE. Our aim is to have a  definite answer to the issue of consistency of the presence of an AdS vacua (presence of a -ve $\Lambda$) in the DE sector with the cosmological observations. We should stress that we do not aim to solve the problem of Hubble tension by adding a -ve $\Lambda$ in the DE sector.

\section{Dark Energy Models with Non-zero Vacua}

Quintessence models, which contain a non zero vacuuam, can be modelled as a rolling scalar field ($\phi$) on top of a cosmological constant $(\Lambda \neq 0)$. Both the field $\phi$ and $\Lambda$ contribute to the total dark energy density of the Universe, $\rho_{de} = \rho_{\phi} + \Lambda$ with the constraint that  $\rho_{de} > 0$ around present time to ensure late time accelerated expansion. 

Instead of picking a particular scalar field with a potential for the quintessence field (\cite{Copeland:2006wr,Tsujikawa:2013fta}),
 we consider two most widely used parametrisations for the equation of state $w_{\phi} = p_{\phi}/\rho_{\phi}$ for the field $\phi$. In one case, $w_{\phi} = w_{0}$ is a constant and it is usually referred as the ``wCDM" model for $\phi$. In another case $w_{\phi} = w_{0} + (1-a)w_{a}$ where $w_{0}$ and $w_{a}$ are constants. This is referred as the ``CPL" model (\cite{Chevallier:2000qy, Linder:2002et}) for $\phi$. The $\rho_{\phi}$ in these two cases scale as:
\begin{eqnarray}
\rho_{\phi} &\sim &a^{-3(1+w_{0})}\nonumber\\
\rho_{\phi} &\sim &a^{-3(1+w_{0} + w_{a})}e^{3w_{a}a}
\end{eqnarray}
Here $a$ is the scale factor in the FRW  metric and we normalize it at present day, $a_{0} =1$. 

With this, for a spatially flat Universe, the FRW equation for the Hubble parameter $H(a)$ is given by
\begin{eqnarray}
    H^{2} &=& H_{0}^{2}[\Omega_{m}a^{-3}+\Omega_{r}a^{-4}+ \Omega_{de}(a)]\nonumber\\
    \Omega_{de}(a) &= &\Omega_{\phi} f(a) + \Omega_{\Lambda}
\label{hubble} 
\end{eqnarray}
where $\Omega_{m}$, $\Omega_{r}$, $\Omega_{\phi}$ and $\Omega_{\Lambda}$ are the density parameters at present for matter (Baryons+CDM), radiation, scalar field $\phi $ and $\Lambda$. The form for $f(a)$ is given by eqn (1) for wCDM or CPL models. As the Universe is spatially flat, we have the constraint $\Omega_{m} + \Omega_{r} + \Omega_{\phi} + \Omega_{\Lambda} = 1$. In our subsequent discussions, the dark energy model with constant $w_{0}$ plus a $\Lambda$, we term as ``wCDMCC" model and the dark energy model with CPL type equation of state plus a $\Lambda$, we term as ``cplCDMCC". In addition to these two specific models with non zero vacuuam, we also consider the usual concordance $\Lambda$CDM model where $\rho_{\phi} = 0$ and $\rho_{de} = \Lambda$.

Let us briefly study the effect of adding a non-zero $\Lambda$ in the DE sector to achieve higher $H_{0}$ keeping the early universe physics unchanged. As we are not interested in modifying the early time cosmology around the recombination period, the sound horizon at recombination ($r_{s}$) is unchanged. Moreover the angular scale of the sound horizon ($\theta$) can be inferred from the peak spacing in the oscillations of CMB temperature anisotropy. It is a very precisely measured quantity, fixed by the latest Planck measurement. $\theta$ is related to $r_{s}$ as $\theta= \frac{r_{s}}{D_{A}}$ where $D_{A}$ is the angular diameter distance to the last scattering surface:
\begin{equation}
    D_{A} \propto \frac{1}{H_{0}(1+z*)}\int_{0}^{z*}\frac{dz}{E(z)},
\end{equation}

\begin{figure}
\centering
{\includegraphics[width=0.48\textwidth,height=8cm]{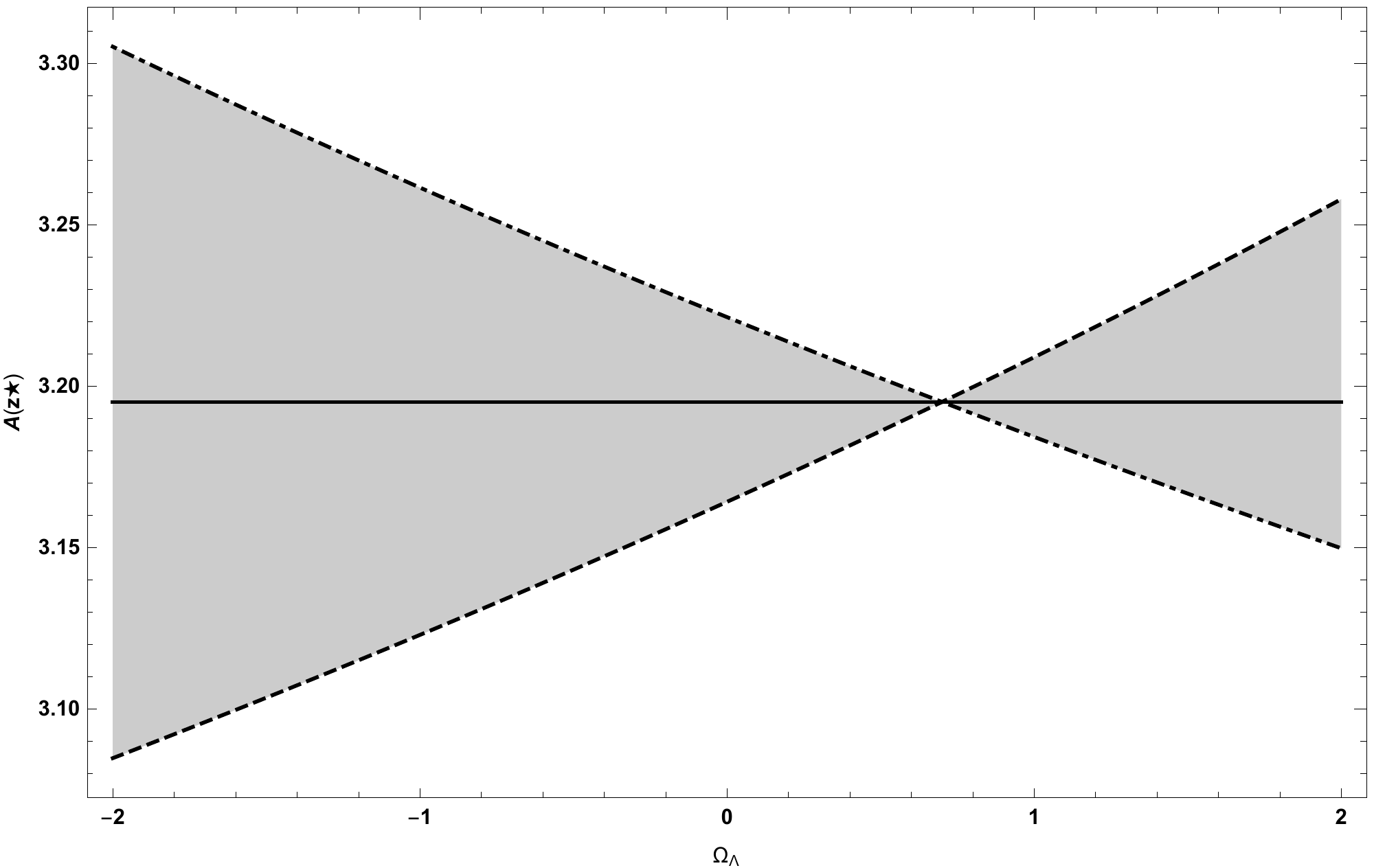}}
\caption{Variation $A(z*)$ vs $\Omega_{\Lambda}$ (see text for definition of $A(z*)$. The solid line is for $
\Lambda$CDM, the Dashed line is for $w= -0.9$ (non-phanatom) and Dot-Dashed line for $w = -1.1$ (phantom) models. We have fixed $\Omega_{m} = 0.3$.  }
\label{A_z}
\end{figure}

\noindent
where $z* \sim 1100$ is the redshift at recombination and $E(z) = [\Omega_{m}(1+z)^{3}+\Omega_{r}(1+z)^{4}+ \Omega_{de}(z)]$ . Given that both $r_{s}$ and $\theta$ are fixed, $D_{A}$ is also fixed. Hence in order to incorporate a larger value of $H_{0}$, keeping $D_{A}$ fixed, the term $A(z*) = \int_{0}^{z*}\frac{dz}{E(z)}$ has to be increased. 

In Figure 1, we show the variation of $A(z*)$ with $\Omega_{\Lambda}$ for wCDMCC model assuming both non-phantom as well as phantom type equation of state for the $\rho_{\phi}$. For comparison, we also show the $\Lambda$CDM result. All the three model behaviours coincide at $\Omega_{\Lambda}$ = 0.7 as expected( we ignore $\Omega_{r}$ and fix $\Omega_{m} =0.3$ for these plots). It is also evident that with $\Omega_{\Lambda} = 0$, we always need phantom type equation of state for the $\rho_{\phi}$. 

We assume $w=-1.1$ and $w=-0.9$ for phantom and non phantom equation of states. For $ -0.9 > w > -1.1 $, all the lines will lie in the shaded region shown the Figure 1 and for $-0.9 < w$ as well as $w < -1.2$, they will lie outside the shaded regions shown in Figure 1.

As one see in Figure 1,  phantom model with both +ve and -ve $\Lambda$ as well as non-phantom model with +ve $\Lambda$ give larger $A(z*)$ compared to $\Lambda$CDM. But the positive deviation from $\Lambda$CDM is larger for phantom models with -ve $\Lambda$. In this regard, we want to mention that same thing can happen for BAO measurements. Usually BAO measurements are related $\frac{r_{d}}{D_{A}}$ where $r_{d}$ is the sound horizon at drag epoch and depends on the early Universe physics. $D_{A}$ is the angular diameter distance to the redshift related to the BAO measurement. For a particular BAO measurement and for fixed $r_{d}$, $D_{A}$ should be fixed and we again need a higher $A(z_{bao})$ for increased $H_{0}$ which can also be obtained in the similar way as mentioned above.
So adding a -ve $\Lambda$ (in other words a -ve vacuum) in the dark energy component, helps to achieve higher $H_{0}$ without changing the sound horizon at recombination ($r_{s}$) or at drag epoch ($r_{d}$).  

\begin{figure}
\centering
{\includegraphics[width=0.48\textwidth]{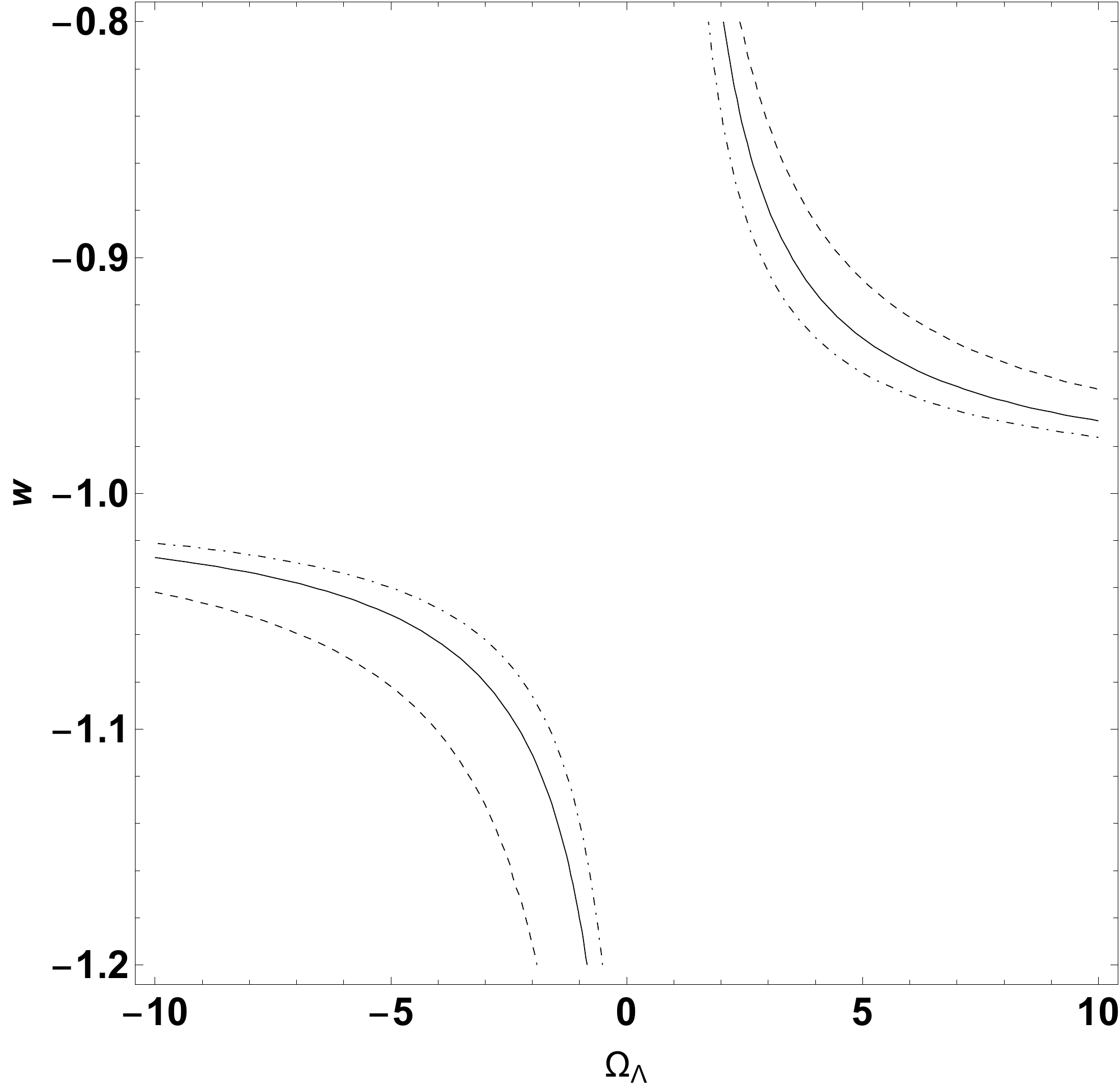}}
\caption{$\Omega_{\Lambda}$vs $w$ as described in the text. Solid line is for $D_{A}$ at $z=0.38$, Dot-Dashed line is for $D_{L}$ at $z=0.5$ and Dotted line is for $D_{A}$ at $z*= 1100$. Please see text for the explanation of these lines.}
\label{Omega_W}
\end{figure}

In Figure 2, we study the same thing in the $w-\Omega_{\Lambda}$ plane. We first consider three observable: $D_{A}$ at last scattering ($z*=1100$), $D_{A}$ at one of the BAO measurements e.g at $z=0.38$ and $D_{L}$ which is the luminosity distance related to SnIa observations at $z=0.5$. We calculate these quantities using $\Lambda$CDM model with $\Omega_{m} = 0.3$ and $H_{0} = 67$ km/s/Mpc. Then in Figure 2, we show the combinations of $w-\Omega_{\Lambda}$
which can reproduce the $\Lambda$CDM results but with $H_{0} = 72$ km/s/Mpc. As one can see from the figure, there are two separate regions: one is phantom models with -ve $\Lambda$ and another is non-phantom models with +ve $\Lambda$. For positive $\Omega_{\Lambda}$, flatness condition plus the positivity of $\rho_{\phi}$ as well as $\rho_{m}$ demand that $\Omega_{\Lambda} \leq 1$. And as one can see from Figure 2, it is difficult to satisfy this constraint with non phantom equation of state. On the, other hand, there is no such constraint for $\Omega_{\Lambda} < 0$. This confirms that adding a -ve $\Lambda$ or a AdS vacuum in the dark energy sector helps in achieving higher $H_{0}$ easily.

In the next sections, we study which region in the $w-\Omega_{\Lambda}$ parameter space as shown in Figure 2, is allowed by the combinations of CMB+BAO+SnIA+R21 data.

\section{Methodology}

In our analysis, we use the most recent and relevant observational results as follows:

\begin{itemize}
    \item CMB data: We consider the recent CMB likelihood by Planck-2018 Legacy release including the lensing likelihood. We use the Planck(2018)- TTTEEE+lowl-TT+lowl-EE+Lensing (\cite{Aghanim:2018eyx,Planck:2019nip,Planck:2018lbu}).
    
    \item Type-Ia Supernova measurement: We consider the Pantheon sample of $1048$ SnIa measurements of luminosity distance (\cite{sn}).
    
    \item BAO observations: We also consider the BAO measurements by SDSS-MGS, BOSS-DR12 as well as 6dFGS collaborations (\cite{BOSS:2016wmc, Beutler, Ross:2014qpa}).
    
    \item Finally we consider the data $H_{0} = 73.30 \pm 1.04$ Km/s/Mpc as measured by SH0ES collaboration (R21) (\cite{Riess:2021arx}).
\end{itemize}

\begin{table*}
\caption{Best fit , Mean and 1$\sigma$ values of the parameters for wCDMCC model.}	
\renewcommand{\arraystretch}{1.5}
	
\begin{tabular}{|c|c|c|c|}
\hline 
Parameters & CMB & CMB+BAO & CMB+BAO+$H_{0}$ \\  
\hline 
{\boldmath$\Omega{}_{m }$} &($0.215$) $0.200^{+0.049}_{-0.068} $ &($0.2956$)$0.304\pm 0.012$ &($0.2793$)$0.2791\pm 0.0065$\\
{\boldmath$H_{0}$(km/s/Mpc)} & ($81.43$)$87.2^{+9.6}_{-16} $ &($69.6$) $68.5^{+1.3}_{-1.5}$&($71.62$)$71.66\pm 0.85$\\	
{\boldmath$r_{d }$} &($147.2$) $147.20\pm 0.38 $ &($147$)$147.17\pm 0.22$&($147$)$147.01\pm 0.22$\\
{\boldmath$\sigma_{8} $} & ($0.936$)$0.970^{+0.085}_{-0.11} $ &($0.8335$)$0.818\pm 0.015$ &($0.8525$)$0.853\pm 0.011$\\
{\boldmath$\tau{}_{re } $} & ($0.05548$)$0.0550\pm 0.0025 $ & ($0.05494$)$0.0549\pm 0.0026$ &($0.05492$)$0.0541\pm 0.0026$\\
{\boldmath$\Omega{}_{\phi }$} &($6.63$) $2.80^{+0.17}_{-2.3}$&($1.584$)$1.86^{+0.46}_{-1.3}$ &($1.539$)$1.58^{+0.16}_{-0.87}$\\
{\boldmath$w_{0 }$} &($-1.035$) $-1.47\pm 0.51$ &($-1.034$)$-1.017^{+0.030}_{-0.015}$ &($-1.061$)$-1.072^{+0.037}_{-0.015}$ \\
{\boldmath$\Omega{}_{\Lambda }$} & ($-5.845$)$-2.00^{+2.5}_{-0.18}$ &($-0.8801$) $-1.17^{+1.3}_{-0.46}$ &($-0.8182$)$-0.86^{+0.87}_{-0.16}$\\
{\boldmath$\Omega{}_{de }$} & ($0.7849$)$0.7999^{+0.068}_{-0.049}$ &($0.7043$) $0.6959^{+0.012}_{-0.012}$&($0.7206$)$0.7208^{+0.0065}_{-0.0065}$\\
\hline

\end{tabular}

\label{table1}
\end{table*}
\renewcommand{\arraystretch}{1}

\begin{table}
\caption{Evidence for wCDMCC model for CMB+BAO+$H_{0}$ data.}	
\renewcommand{\arraystretch}{1.5}
	
\begin{tabular}{|c||c|c|c|c|c|}

\hline

Model&{\boldmath$\chi^{2} $} & {\boldmath$AIC$} & {\boldmath$ln(z)$} &{\boldmath$\Delta$AIC}&{\boldmath$\Delta$$ln(z)$}\\
\hline
$\Lambda$CDM&2803.04 &2869.04 & -1427.829&0&0\\

wCDMCC&2782.40 &2852.40 & -1420.259&-16.64&7.561\\	
\hline
\end{tabular}

\label{table2}
\end{table}
\renewcommand{\arraystretch}{1}

\begin{table}
\caption{Best fit , Mean and $\sigma$ values of the parameters for wCDMCC model.}	
\renewcommand{\arraystretch}{1.5}
	
\begin{tabular}{|c|c|c|}
\hline
Parameters & CMB+SN & CMB+SN+BAO \\  
\hline 

{\boldmath$\Omega{}_{m }$} &($0.3021$)$0.305\pm 0.010$ &($0.3045$)$0.3053\pm 0.0077$\\
{\boldmath$H_{0}$(km/s/Mpc)} &($68.77$) $68.5\pm 1.1$&($68.64$)$68.39^{+0.79}_{-0.92}$\\
{\boldmath$r_{d }$} &($147$) $147.12\pm 0.22$&($146.9$)$147.14\pm 0.21$\\
{\boldmath$\sigma_{8} $} &($0.8207$)$0.821\pm 0.011$ &($0.8252$)$0.8185^{+0.0094}_{-0.011} $\\
{\boldmath$\tau{}_{re } $} &  ($0.05374$)$0.0543\pm 0.0026$ &($0.05455$) $0.0547\pm 0.0025$\\
{\boldmath$\Omega{}_{\phi }$} &($0.8285$)$1.91^{+0.22}_{-1.3}$ &($2.703$) $1.52^{+0.16}_{-0.91}$\\
{\boldmath$w_{0 }$} &($-1.031$)$-1.016^{+0.020}_{-0.012}$ &($-1.011$)$-1.017^{+0.023}_{-0.012}$ \\
{\boldmath$\Omega{}_{\Lambda }$} &($-0.1306$) $-1.21^{+1.3}_{-0.22}$ &($-2.008$)$-0.83^{+0.92}_{-0.16}$\\
{\boldmath$\Omega{}_{de }$} &($0.6978$)$0.6949^{+0.010}_{-0.010}$ &($0.6954$)$0.6946^{+0.0077}_{-0.0077}$ \\
{\boldmath$M_{B}$}&($-19.39$) $-19.402\pm 0.023$ &($-19.4$)$-19.404\pm 0.018$  \\

\hline

\label{table3}
\end{tabular}
\end{table}
\renewcommand{\arraystretch}{1}

\begin{table*}
\caption{Constraints (Best fit, mean and $1\sigma$ error bar) on cosmological and model parameters for the combined data CMB+BAO+Pantheon+R21.}
\centering
\renewcommand{\arraystretch}{1.5}
\begin{tabular}{|c|c|c|c|}
\hline 
Parameters & $\Lambda$CDM&wCDMCC&cplCDMCC \\ 
\hline
{\boldmath$\Omega_{m}$}&($0.3033$)$0.3043_{-0.0042}^{+0.0039}$&($0.287$)$0.288_{-0.0071}^{+0.0063}$&($0.292$)$0.289_{-0.0061}^{+0.0055}$\\ 
{\boldmath$H_{0}(km/s/Mpc)$}&($68.32$)$68.2_{-0.32}^{+0.31}$&($70.34$)$70.32_{-0.77}^{+0.82}$&($69.89$)$70.25_{-0.69}^{+0.67}$\\
{\boldmath$r_{d}$}&($147.2$)$147.3_{-0.22}^{+0.2}$&($147.3$)$147.1_{-0.22}^{+0.21}$&($147.2$)$147.1_{-0.21}^{+0.2}$ \\
{\boldmath$\sigma{}_8$}&($0.804$)$0.805_{-0.0026}^{+0.0026}$&($0.833$)$0.834_{-0.0097}^{+0.0096}$&($0.835$)$0.835_{-0.0088}^{+0.0086}$\\
{\boldmath$\tau{}_{re }$} & ($0.0548$)$0.0556_{-0.0026}^{+0.0025}$&($0.0554$)$0.0547_{-0.0025}^{+0.0028}$&($0.0542$)$0.0547_{-0.0027}^{+0.0026}$\\
{\boldmath$\Omega{}_{\phi }$}&--&$(3.191)$ $2.504_{-1.9}^{+0.28}$&$(1.206)$ $1.492_{-0.84}^{+0.025}$\\
{\boldmath$\Omega{}_{\Lambda{}}$}&--&($-2.479$)$-1.792_{-0.29}^{+1.90}$&($-0.498$)$-0.781_{-0.03}^{+0.85}$\\
{\boldmath$\Omega{}_{de }$} &($0.6966$)$0.6956_{-0.0039}^{+0.0042}$ &($0.7129$)$0.7119_{-0.0063}^{+0.0071}$&($0.7079$)$0.7109_{-0.0055}^{+0.0061}$\\
{\boldmath$w_{0}$}&-&($-1.02$)$-1.04_{-0.013}^{+0.035}$&($-1.02$)$-1.03_{-0.039}^{+0.051}$\\
{\boldmath$w_{a}$}&-&-&($-0.12$)$-0.10_{-0.14}^{+0.20}$\\
{\boldmath{$M_{B}$}}&{($-19.401$)$-19.404^{+0.00929}_{-0.00921}$}&{($-19.362$)$-19.363^{+0.0150}_{-0.0151}$}&{($-19.366$)$-19.360^{+0.0166}_{-0.0168}$}\\
\hline 
\end{tabular}

\label{table4}
\end{table*}
\renewcommand{\arraystretch}{1}

\begin{table}
\caption{Evidences of different models for all data.}	
\renewcommand{\arraystretch}{1.5}
	
\begin{tabular}{|c|c|c|c|c|c|}
\hline 
 
Model&{\boldmath$\chi^{2} $} & {\boldmath$AIC$} & {\boldmath$ln(z)$} &{\boldmath$\Delta$AIC}&{\boldmath$\Delta$$ln(z)$}\\
\hline
$\Lambda$CDM&3835.73 &3901.73 & -1944.75&0&0\\	
wCDMCC&3823.58 &3893.58 & -1940.96&-8.15&+3.79\\
cplCDMCC&3824.29 &3896.29 & -1942.82&-5.44&+1.93\\

\hline
\end{tabular}

 \label{table5}
\end{table}
\renewcommand{\arraystretch}{1}

We use publicly available CLASS code (\cite{class}) to compute the CMB temperature anisotropy spectra for the Planck likelihood. To constraint the cosmological models using the above set of observational data, we use the publicly available MCMC sampling algorithm MontePython (\cite{Brinckmann:2018cvx,Audren:2012wb}) to generate the chains. We adopted the Gelman-Ruben convergence criteria $R-1$ $\mathcal{<} 0.02$ and subsequently use the GetDist code (\cite{getdist}) to analyse the chains.

In our analysis, we only change the dark energy sector by incorporating non zero vacua ($\Lambda \neq 0$) for the quintessence field and we do not vary the primordial power spectrum (PPS) due to early inflationary era and hence keep the parameters $n_{s}$ and $A_{s}$ fixed at values as obtained by Planck-2018 (\cite{Aghanim:2018eyx}). The main cosmological parameters are $\Omega_{m}$, $r_{d}$, $H_{0}$, $\sigma_{8}$ and $\tau_{re}$ which are common for all three models. In addition to this, wCDMCC has two extra parameters $w_{0}$ and $\Omega_{\Lambda}$ and cplCDMCC has three extra parameters $w_{0}$, $w_{a}$ and $\Omega_{\Lambda}$.

Regarding the priors for different parameters, we use the same priors already incorporated in the MontePython Code(\cite{Brinckmann:2018cvx,Audren:2012wb}) while using the Planck 2018 likelihood for $\Lambda$CDM, WCDM and CPL models.
For the extra parameter due to the presence of $\Lambda$, we use the uniform prior for $\Omega_{\phi}$, $[0.6,4.5]$ and consider $\Omega_{\Lambda}$ as a derived parameter.Note that we have not assumed any prior for $\Lambda$ to be negative as we need to study the $\Omega_{\Lambda} -w$ parameter space as shown in Figure 2, which theoretically allows both positive and negative $\Omega_{\Lambda}$. The lower limit for $\Omega_{\phi}$ prior is chosen such that we always have sufficient scalar field energy density to accelerate the Universe. We want to stress that the acceleration of the Universe is due to to slow rolling of the scalar field over its potential and it is sourced primarily by $\Omega_{\phi}$ which is always positive. The $\Omega_{\Lambda}$ contribution in $\Omega_{de}$ as in eqn (2), is due to non zero minimum for the scalar field potential which can be both positive or negative due to nonzero positive or negative minimum of the scalar field potential. Also due to the assumption of spatial flatness, $\Omega_{de} = 1 - \Omega_{m} -\Omega_{r}$ and hence is always positive.

\section{Results}

To start with, we use the simplest wCDMCC model and study the effect of adding a non-zero vacuum in the scalar field dark energy through the addition of $\Omega_{\Lambda}$ and its effect on the measured value of the parameter $H_{0}$. First we want to see whether for wCDMCC model, the constraints on $H_{0}$ from CMB and CMB+BAO are consistent with the local $H_{0}$ measurement by R21. This is important for further adding the R21 $H_{0}$ data with CMB+BAO data.

As shown in Table (\ref{table1}), the CMB only constraint on $H_{0}$ is consistent with R21 measurement for $H_{0}$. Moreover the best fit value for $H_{0}$ from CMB+BAO data combination is also consistent at around $2\sigma$ with R21 constraint on $H_{0}$. This is already a substantial improvement from the $\Lambda$CDM model. Also for both CMB and CMB+BAO combinations, the best fit values for $\Omega_{\Lambda}$ is -ve, confirming the consistency of the non zero AdS vacuum for the dark energy. These allow us to add the R21 data for $H_{0}$ together with the CMB+BAO combination which is shown in the last column of Table (\ref{table1}). It is evident that CMB+BAO+${H_{0}}$ data is consistent with -ve $\Omega_{\Lambda}$ showing that dark energy with non zero AdS vacuum is fully consistent with cosmological observations. Although we have a higher $H_{0}$, but still the constraint on $r_{d}$ is similar to  Planck-2018 constraint for $\Lambda$CDM model showing that the early time physics is unaffected.

To compare the wCDMCC model with $\Lambda$CDM for CMB+BAO+$H_{0}$ data combination, we adopt the model comparison criteria based on the Akaike Information Criteria (AIC) (\cite{aic}) which defined as

\begin{equation}
    AIC = \chi^2_{min} + 2k,
\end{equation}

\noindent where $\chi^2_{min}$ is the $\chi^2$ value for the best fit parameters of the model and $k$ is the number of model parameters. Model with lower AIC is preferred over model with higher AIC.  We also compute the ``Bayesian Evidence" ({$\mathcal Z$}) or marginalized global likelihood for each model. We use the publicly available MCEvidence package (\cite{Heavens:2017afc}) for this purpose. The quantity $\log {\mathcal Z}$ is an estimator to compare different models. The results are shown in Table (\ref{table2}). As shown in the table, for both the criteria, wCDMCC is decisively favoured over $\Lambda$CDM model for CMB+BAO+$H_{0}$ data combination. This confirms that the local R21 measurement of $H_{0}$ when combined with CMB and BAO observations, does signal the existence of a non zero AdS vacuum (existence of a non zero -ve $\Lambda$) in the dark energy sector.

Next we do similar analysis while adding the SnIa Pantheon data with CMB. The results are shown in Table (\ref{table3}). For CMB+SN, the best fit value for $H_{0}$ is consistent with Local R21 measurement at around $3\sigma$ which is not as good as for CMB+BAO, but still there is substantial improvement from $\Lambda$CDM which is at $5\sigma$ tension from R21 result. We also show the constraints for CMB+BAO+SN data combination and the constraint on $H_{0}$ is similar to CMB+SN combination. For both combination of data, the best fit value for $\Omega_{\Lambda}$ is -ve, confirming that the CMB+BAO+SN also allow non zero AdS vacuum for the dark energy (in other words, presence of a non zero negative cosmological constant in the dark energy sector).

Going further, we finally consider all the data e.g CMB+BAO+SN+$H_{0}$. It has been discussed in the literature that Pantheon SnIa data and local $H_{0}$ measurement may not be considered together due to the dependency of the local $H_{0}$ measurement on the Pantheon measurements at redshifts within the Hubble flow (\cite{Camarena_2021}). This is more important for models where the late time effects to increase $H_{0}$ are more prominent for $z < 0.02$ (\cite{Abdalla_2022}). Example for such models is hockey-stick dark energy model (\cite{Camarena_2021}) . In our case, we do not have such late transition in dark energy behaviour. Hence we consider the full data combination to see how the constraints get affected. The results are shown on Table (\ref{table4}). In this case, we also consider the cplCDMCC model. For wCDMCC case, the results are similar to CMB+BAO+$H_{0}$ combination except that more negative values for $\Omega_{\Lambda}$ is allowed. The equation of state for the dark energy is slightly phantom. Moreover, wCDMCC model allows larger contribution from negative $\Lambda$ in the energy budget of the Universe compared to cplCDMCC model. These results are similar to what obtained earlier by Visinelli et al (\cite{ads4}). Interestingly, with the inclusion of CMB+lensing data as well as the new R21 measurement for $H_{0}$, we get much stronger constraints on the lower limit of $\Omega_{\Lambda}$ compared to the results by Visinelli et al (\cite{ads4}).Moreover the constraints on $M_{B}$ for $\Lambda$CDM model is $4.14\sigma$ away from the $M_B$ prior corresponding to the SN measurements from SH0ES (\cite{Camarena_2021}) whereas for wCDMCC, it is less than $3\sigma$ away from the SH0ES prior on $M_{B}$ showing improvement over $\Lambda$CDM.(See (\cite{Nunes_2021}) for similar results related to wCDM and interacting dark energy models.)

In Table (\ref{table5}), we show the $\chi^2_{min}$, AIC and $Ln(Z)$ values for different models for ``ALL" data considered. As one can see, when compared with $\Lambda$CDM, both $\Delta AIC$ and $\Delta Ln (Z)$ decrease compared to CMB+BAO+$H_{0}$ combination. This is due to the addition of the SN data. But dark energy models with non zero AdS vacuum still fare substantially better than $\Lambda$CDM model.


%

\begin{figure}
\centering
{\includegraphics[width=0.48\textwidth]{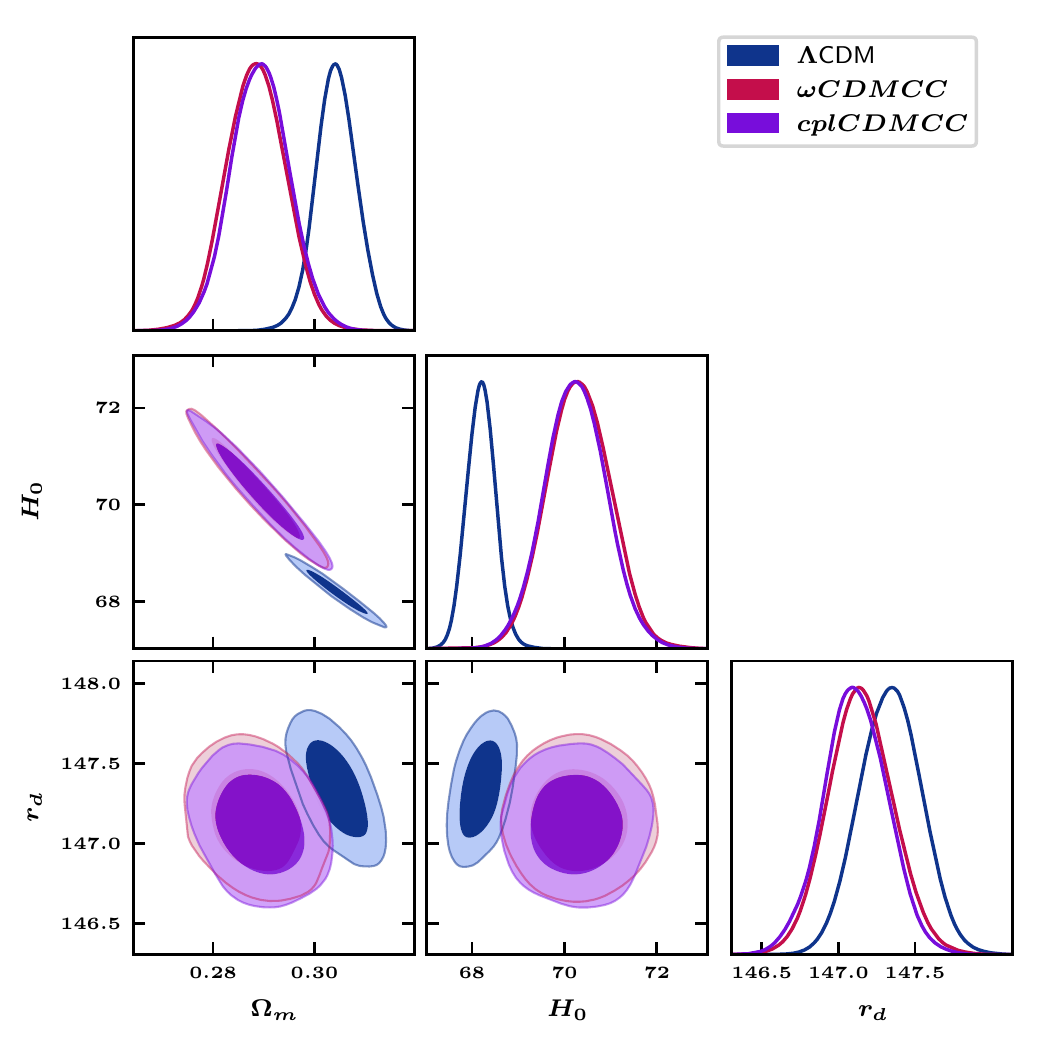}}
\caption{Combined likelihood and contour plots ($68\%$ and $95\%$) of different models for CMB+Pantheon+BAO+R21 data combination. }
\label{contour1}
\end{figure}

\begin{figure}
\centering
{\includegraphics[width=0.48\textwidth]{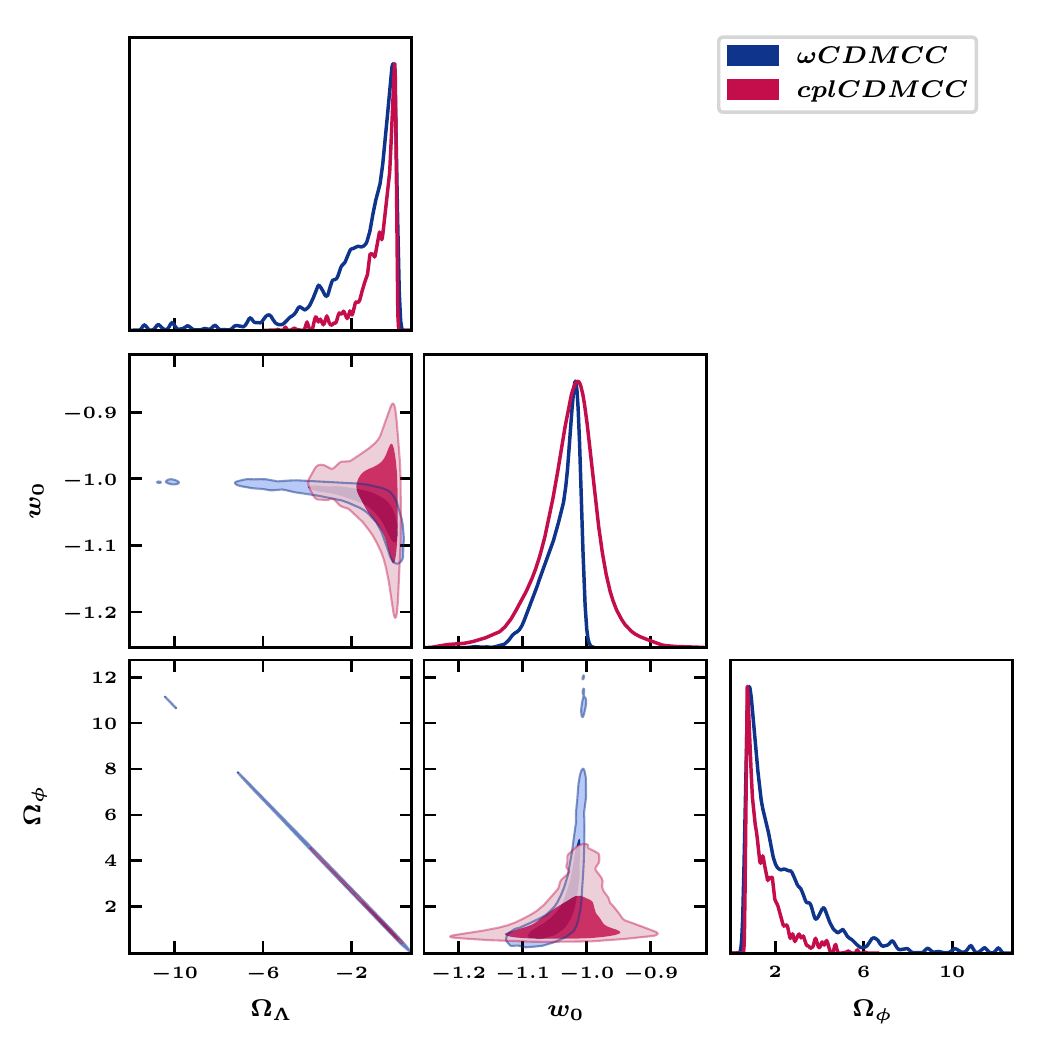}}
\caption{Combined likelihood and contour plots ($68\%$ and $95\%$) of different models for CMB+Pantheon+BAO+R21 data combination.}
\label{contour2}

\end{figure}

For completeness, in Figure (\ref{contour1}) and Figure (\ref{contour2}) we show the likelihood and contour plots for few relevant parameters in different models. It is interesting to note from Figure (\ref{contour2}) that for wCDMCC, the shape of confidence contour in the $w_{0}-\Omega_{\Lambda}$ plane is consistent with what we show in Figure 2 justifying our physical argument in section 2.



\begin{figure}
\centering
{\includegraphics[width=0.48\textwidth,height=10cm]{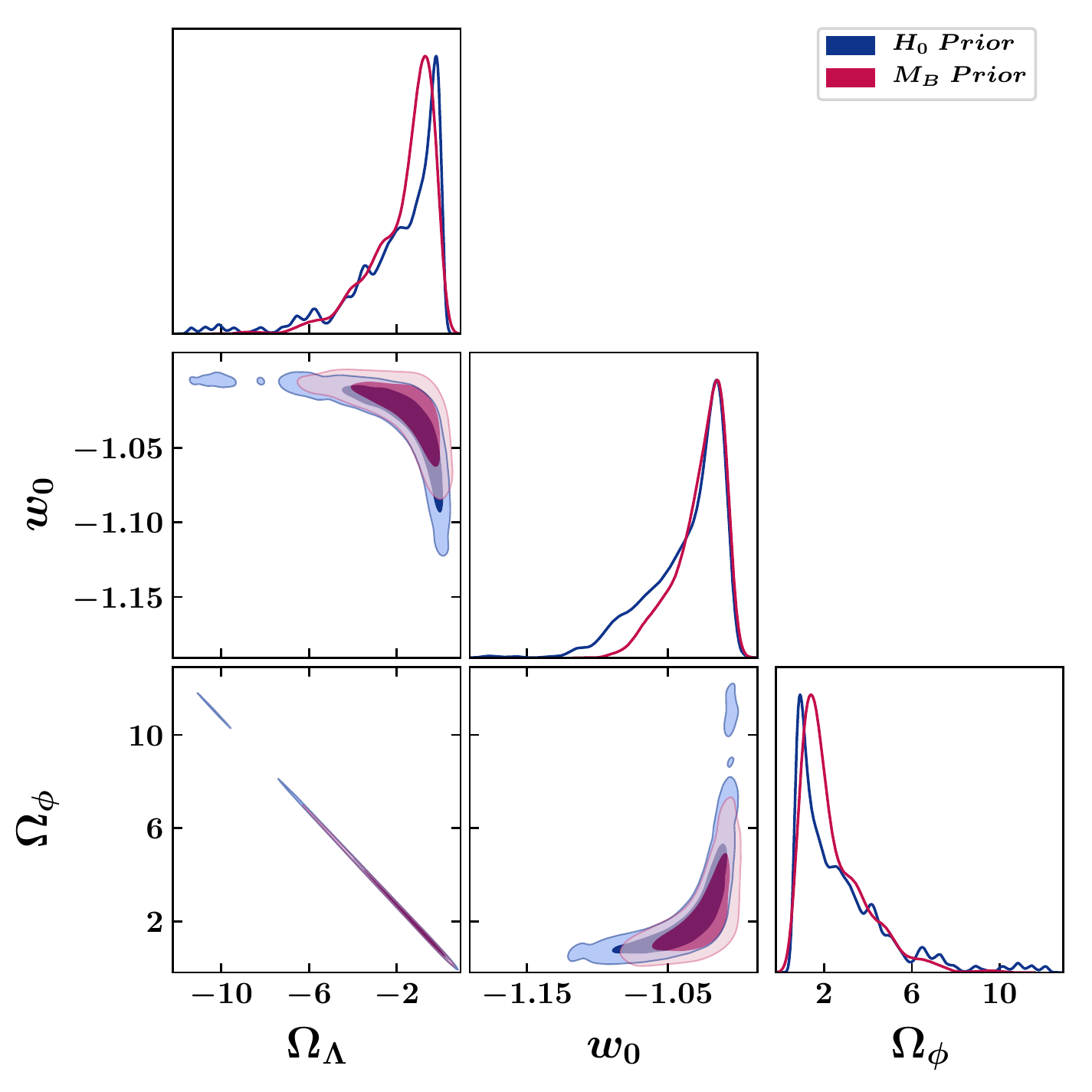}}
\caption{Same as Figure 4 but with $M_{B}$ prior.}
\label{mbprior}

\end{figure}

Finally we also use the absolute magnitude $M_{B}$ prior instead of local $H_{0}$ prior in all data combination (\cite{Camarena_2021}). The constraint  on $H_{0}$ for wCDMCC case is $H_{0} = 69.70 ^{+0.71}_{-0.80}$ km/s/Mpc. This is consistent with constraints on $H_{0}$ while using $H_{0}$ prior. In Figure (\ref{mbprior}), we show the comparison between $H_{0}$ and $M_{B}$ prior for constraints on model parameters e.g $\Omega_{\Lambda}$, $w_{0}$ and $\Omega_{\phi}$. It is evident that the results are almost identical with $M_{B}$ prior marginally favours larger negative values for $\Omega_{\Lambda}$.

The $\chi_{min}^2$, AIC and Bayesian Evidence numbers for $\Lambda$CDM and wCDMCC models while using $M_{B}$ prior are also shown in Table (\ref{table5}). The estimators like $\chi_{min}^{2}$ or AIC are similar to that for $H_{0}$ prior and hence the conclusions with $H_{0}$ prior still hold with $M_{B}$ prior. In terms of the Bayesian Evidence, $\Lambda$CDM and wCDMCC are equally consistent and none is preferred over the other. This also confirms that models with -ve $\Lambda$ (with AdS vacuum) are consistent with present cosmological observations with $M_{B}$ prior as well.
\begin{table}
\caption{$\chi_{min}^2$, $\Delta$ AIC and Bayesian Evidences for $\Lambda$CDM and wCDMCC for CMB+Lensing+Pantheon+BAO+$M_{B}$ data combination. Here $\Delta$AIC is saame as defined in Table 3.}
\centering
\begin{tabular}{|l|c|c|c|}  %
	\hline
	Model & $\chi_{min}^{2}$ & $\Delta$AIC & {$\log(\mathcal Z)$} \\
	\hline
	 $\Lambda$CDM  & 3833 & --& -1941.97  \\
	  \hline
	  wCDMCC  & 3824 & -5.0 &  -1942.04 \\
	  
	  \hline 
	  
\end{tabular}

\label{table6}
\end{table}

\section{Few Remarks About the Model}

\begin{figure}
\centering
{\includegraphics[width=0.48\textwidth,height=10cm]{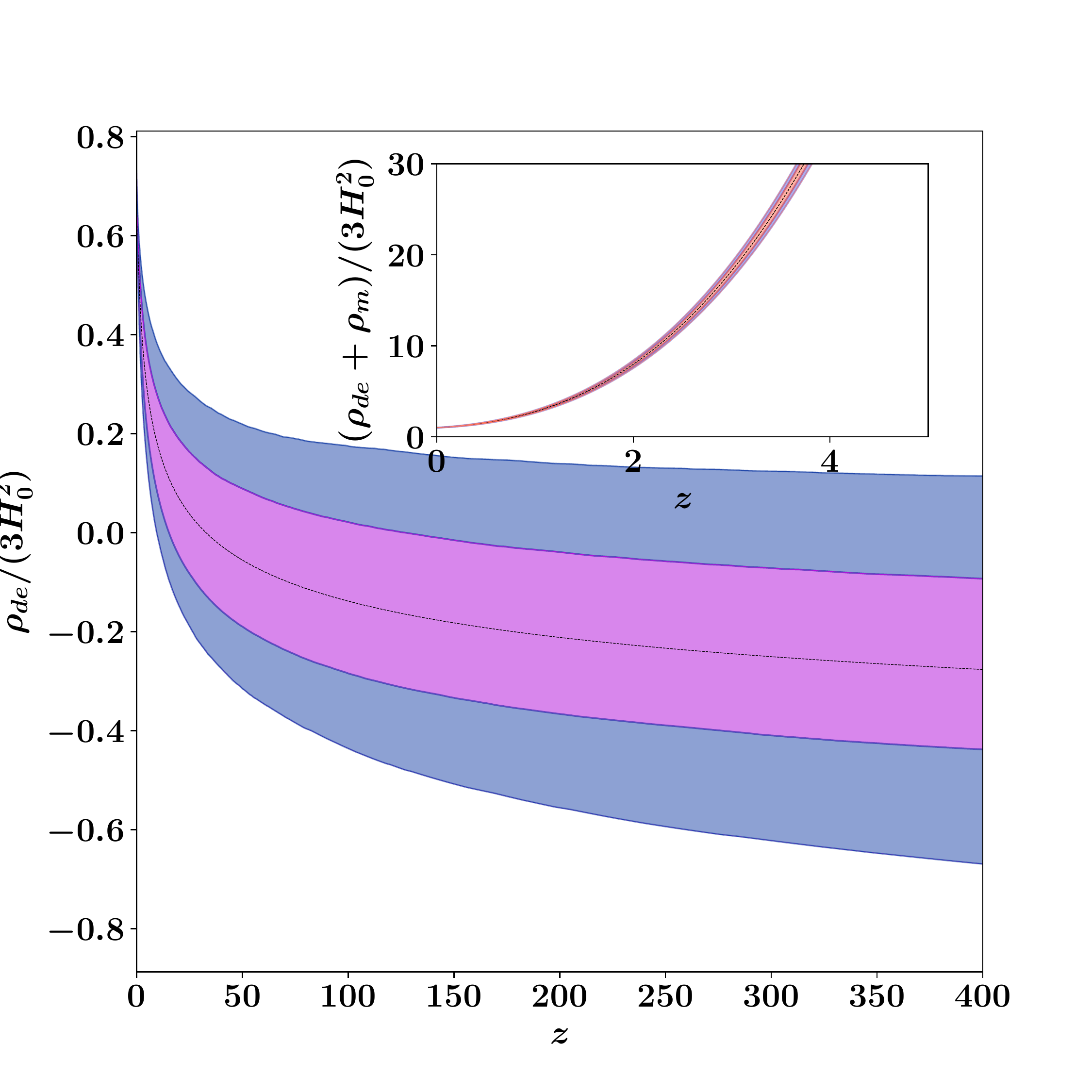}}
\caption{Reconstruction of $\rho_{de}$/3$H_{0}^{2}$ and ($\rho_{de}$+$\rho_{m}$)/3$H_{0}^{2}$. }
\label{de_recr}

\end{figure}

As discussed in section 2, we are considering dark energy models where the dark energy scalar field has non zero vacuum. As shown in the results, the data prefers that this vacuum should be negative showing the existence of non-zero negative $\Lambda$ in the dark energy sector. The total dark energy density $\Omega_{de} = \Omega_{\phi} + \Omega_{\Lambda}$ has the contribution from the scalar field dark energy density which is $\Omega_{\phi}$ as well as from the negative $\Lambda$  which is $\Omega_{\Lambda}$. This dark energy density $\Omega_{de}$ is positive at present as shown in Table (\ref{table1}),(\ref{table3}) and (\ref{table4}). In Figure (\ref{de_recr}), we also show the reconstructed evolution for $\rho_{de}$ and $\rho_{de} + \rho_{m}$ for the CMB+BAO+$H_{0}$ combination for wCDMCC model. As one sees, the $\rho_{de}$ settles to a small negative value in the past and then increases to a positive value in late times to give the necessary contribution for the dark energy. This negative $\rho_{de}$ behaviour in the past is consistent with earlier studies by Bonilla et al (\cite{Bonilla_2021}).But the total contribution $\rho_{m} + \rho_{de}$ is always positive ruling out any possible collapsing scenario in the past. 

\noindent
The behaviour for $\rho_{de}$ can also lead to another interesting feature with models with AdS vacuum as considered here. Let us consider the conservation equation for $\rho_{de}$: $\rho_{de}^{\prime} = 3(1+w_{de})\rho_{de}/(1+z)$ where `prime' denotes the differentiation with respect to the redshift $z$. As shown in Figure (\ref{de_recr}), $\rho_{de}$ is a decreasing function with redshift $z$. Hence $\rho_{de}^{\prime}< 0$. In the early time, $\rho_{de} < 0$, hence $(1+w_{de}) > 0$ resulting a non-phantom equation of state for dark energy sector whereas at late times, $\rho_{de} > 0$; hence $(1+w_{de}) < 0$ resulting a phantom equation of state for the dark energy sector. This shows that the dark energy sector has a non-phantom to phantom transition as the Universe evolves.Hence adding a negative cosmological constant in the dark energy sector can result phantom crossing. This may be the simplest way to incorporate phantom crossing in the dark energy equation of state.

 One should note that ours is a toy model for dark energy where we model $\omega_{\phi}$ using  simple equation of state parametrization. This results a phantom equation of state for the $\rho_{de}$ as constrained by the observational data. One can have more realistic scenario constructed under string theory   (\cite{ads4}) where there may be multiple light bosonic degrees of freedom on top of a stable AdS vacua (\cite{2005PhLB..608..177G, 2005PhRvD..71d7301H, 2010PhRvD..81l3523S}). Example of such scenario is string axiverse where we have multiple moduli fields related to the shapes and sizes of the extra dimensions (\cite{2006JHEP...06..051S, 2010PhRvD..81l3530A, 2012JHEP...10..146C, 2017PhRvD..96b3013V}). The dynamics of these multiple scalar fields can give rise to an effective phantom equation of state for the DE sector that results the $\rho_{de}$ behaviour as shown in Figure (\ref{de_recr}). But as the fields settle at the AdS ground state in future, the Universe will end up in a asymptotically AdS phase. The future Universe for such a scenario with negative energy density in future has recently been studied in an interesting paper by Andrei, Ijjas and Steinhardt (\cite{Andrei_2022}).
 
\section{Conclusions}

Although consistent with most of the cosmological observations, $\Lambda$CDM model has always been a challenge for theoreticians because of the presence of a dS vacua. It is difficult to construct in string theory and the recent Swampland conjecture put it in tight spot. But with current $5\sigma$ inconsistency between the local measurement of $H_{0}$ by SH0ES (R21) and that by Planck using CMB, $\Lambda$CDM model seems to fall from its preferred position in terms of observational consistency and this has opened up the opportunity to revisit the quintessence models and to consider DE models with AdS vacua which are more natural in string theory. As already mentioned in the Introduction, goal of this study is not solve the $H_{0}$ tension with the introduction of AdS vacua in the dark energy sector rather to see whether dark energy with AdS vacua is consistent with cosmological observation including the local R21 measurement of $H_{0}$.

In our study, we parametrize the quintessence models with two most widely used dark energy representation in literature and add a non zero vacua (in terms of a $\Lambda$ which can be either +ve or -ve) and use the CMB likelihood, the latest $H_{0}$ measurement (R21) as well as the Pantheon and BAO data to constrain the parameter space as well as compare such models with $\Lambda$CDM.

For wCDMCC, both CMB and CMB+BAO data result substantial increase in the constraints for $H_{0}$, decreasing the tension with the local R21 measurement to around $2\sigma$. This allows us to add the local measurement of $H_{0}$ (R21) to CMB+BAO. The combined CMB+BAO+$H_{0}$ data show that wCDMCCC model is decisively favoured over $\Lambda$CDM confirming the consistency of the presence of AdS vacua (in terms of -ve $\Lambda$) in the dark energy sector with the cosmological data. Interestingly the constraint on $r_{d}$ parameter is completely consistent with the Planck-2018 constraint for $\Lambda$CDM showing the early Universe physics is unaffected.

Adding Supernova Pantheon data with CMB, we get $H_{0}$ constraint that is within $3\sigma$ from the local R21 result. This is not as good as result for the CMB+BAO combination but still much better than the $5\sigma$ tension in $\Lambda$CDM model. With this, we consider the full data combination, CMB+BAO+$H_{0}$+SN and show that models with AdS vacuum for dark energy sector is consistent with the cosmological observations and are favoured over $\Lambda$CDM model.
 
In 1998, the SnIa observations surprised us with the confirmation of an accelerating Universe and the presence of DE. Again, after two decades, low redshifts SnIa give us the $H_{0}$ measurement that is in $5\sigma$ conflict with Planck results assuming the $\Lambda$CDM model, the ``holy grail" for cosmologists. This has the potential to reveal another surprising aspect of our Universe. Can it be the presence of a -ve $\Lambda$ in our Universe? We show that this may indeed be true. This can have far reaching implications in dark energy model building, especially  while addressing the different cosmological tensions in the realm of string theory. Our results can be potentially the first cosmological signature of an aspect of string theory, namely the existence of a negative $\Lambda$. Moreover as the Universe is asymptotically AdS, one can have interesting consequences related to the future light-cone structure of the Universe. This can open up many interesting directions for future research.

\section*{Acknowledgements}

The authors thank Dhiraj Hazra, Sunny Vagnozzi, Eleonora de Valentino, Akash Garg, Gaurav Goswami and Ruchika for useful discussions and helpful suggestions. AAS acknowledges the funding from SERB, Govt of India under the research grant no: CRG/2020/004347. SAA is funded by UGC non-NET Fellowship scheme. The authors also acknowledge the use of High Performance Computing facility Pegasus at IUCAA, Pune, India.

\section*{Data Availability}

The data that support the findings of this study are available upon reasonable request.



\bibliographystyle{mnras}
\bibliography{biblio} 





\bsp	
\label{lastpage}
\end{document}